**Scaling laws governing the roughness of the swash edge line**

Ed. Bormashenko[a,b]*, A. Musin,[a,b] R. Grynuov[a],

[a]Department of Physics, Ariel University, P.O.B. 3, Ariel 40700, Israel.[a]
[b]Department of Chemical Engineering and Biotechnology, Ariel University, P.O.B. 3, Ariel 40700, Israel

The physics of swash i.e. a layer of water that washes up on the beach after an incoming wave has broken is complicated and intriguing. It includes perplexed hydrodynamic and sediment transport events [1–4]. In our paper we address to the roughness of the moving swash boundary at which a beach, water and air meet. We treat the behavior of this boundary as an interfacial phenomenon, without going into details of formation of edge waves and beach cusps, covered broadly in literature [1–4]. This "crude" approach turns out to be productive and revealing the resemblance of the swash line with a broad diversity of effects arising from the random pinning of moving boundaries.

Swash consists of two phases: uprush (onshore flow) and backwash (offshore flow). We concentrate mainly on backwash (offshore flow), when water recedes along the beach face. The receding water front is pinned by randomly distributed obstacles, produced by the beach and constituting the roughness of the swash edge line. It is plausible to suggest that the fine structure of the swash edge line results from the combined action of the pinning force applied by defects of the beach and elasticity of distorted swash boundary. The problem is common for a number of physical systems comprising interfaces moving in the disordered medium as, for example, self-affine crack propagation [5–6], wet front propagation and movement of interfaces between immiscible liquids in porous media [7–9], height distribution of growing surfaces [10, 11], growth of magnetic domain walls in disordered media [12]. However, the most investigated situation of interfaces moving over disordered system of obstacles is the displacement of triple (three phase) line, at which water, vapor and rough (or chemically heterogeneous) surface meet [13–17].

The shape of triple line disturbed by the presence of random heterogeneities on the solid surface may be derived by solution of the Laplace equation with the boundary conditions of the Young equation. With some simplifications it is given by Shanahan [18] and de Gennes [19]. The simplest quantitative characteristic of interface is its width $w$, defined as the root-mean-square fluctuation around average position. For rough interfaces scaling with size of the system $L$ is observed in the form

$$w(L) \propto L^{\zeta}, \qquad (1)$$

meaning that the average width $w$ grows as the size $L$ is increased. The concept of scaling introduced in the field of interfaces by Family and Vicsek [20, 21] gave a simple framework for classifying interfaces in various processes. Scaling laws describing the shape of the triple line were studied in many works [13–16]. Constant value of roughness exponent $\zeta$ in a given physical system was proposed to express some universal law and gave rise to different models. Exponent values 1/2 and 1/3 (respectively, for fluctuations smaller and larger than the length scale of defects) were calculated by Robbins and Joanny [23] and Joanny and de Gennes [2] according to the model of long-range elastic forces and supposed to be universal for describing contact lines on surfaces with disordered defects. These values indeed were observed by Rolley *et al*. [14] in the experiments with superfluid helium-4 on a cesium substrate with random disorder. Values of about 0.5 were measured for water meniscus on glass plate covered with random spots of chromium [15]. On the other hand, values of roughness exponent measured in other experiments differed from the given theoretical predictions. The values in the region 0.82–0.87 were measured for water and 0.77–0.86 for hexadecane on chemically degraded fluorinated polymer surface [13]. Experiments performed with helium-4 deposited on a strongly disordered cesium substrate supplied the roughness exponent $\zeta = 0.56 \pm 0.03$ [24]. The value of $\zeta = 0.63$ appeared in a number of different models, based on directed percolation [25]. This value was obtained for a wet front propagation in porous media [8], for the crack propagation fronts in plastics [26, 27].

In our recent work [28] we measured the roughness exponent of triple line of water droplet deposited on the porous polymer (polycarbonate) substrate in the situation of the Cassie wetting and received the values 0.60±0.05 for the receding and 0.63±0.02 for the advancing triple. It was plausible to suppose that exponent value $\zeta = 0.6$–0.63 is universal for the processes when mobile water interface is stopped by randomly distributed pinning sites, forming so called capillary fringes [4, 19]. The backwash phase of swash fulfils these demands, whatever is the nature of a beach.

Thus, the value of the scaling exponent is expected to be the same or at least close to the value established for triple lines. We checked this hypothesis experimentally.

Several hundreds of pictures of a water front and contact line were taken at the high Mediterranean Sea coast in the fair and calm day (wind speed about 2 m/s), see Fig. 1. Two states, uprush (onshore flow) and backwash (offshore flow), presenting advancing and receding contact lines, may be distinguished, and all images were classified according to these two cases.

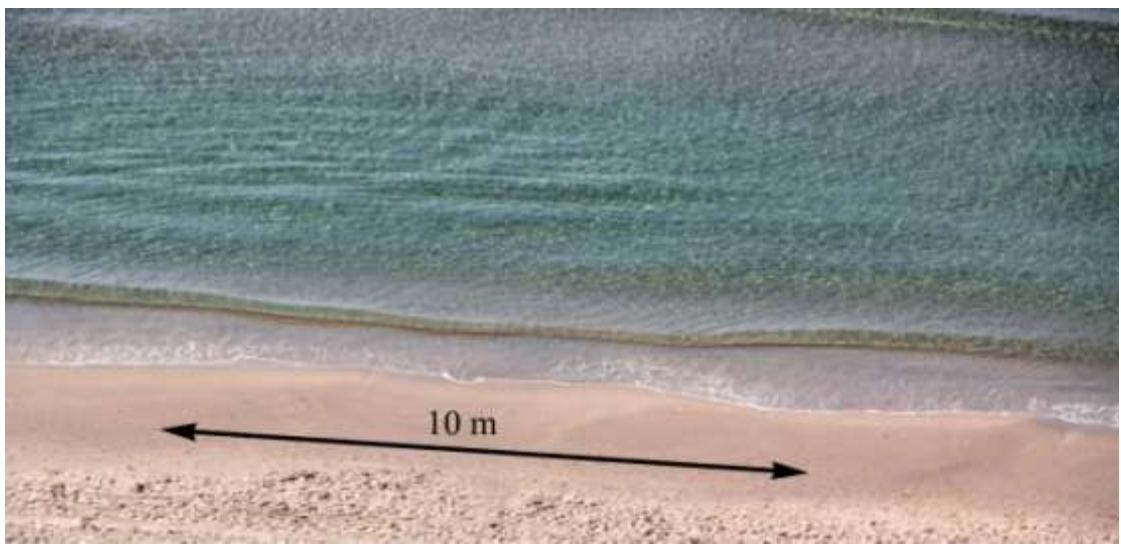

FIG. 1. Propagation of water front and receding contact line.

Position of contact lines on images were digitized according to the method described earlier [28] and characterized by coordinates $x$ along a line and $y$ in the perpendicular direction (Fig. 2).

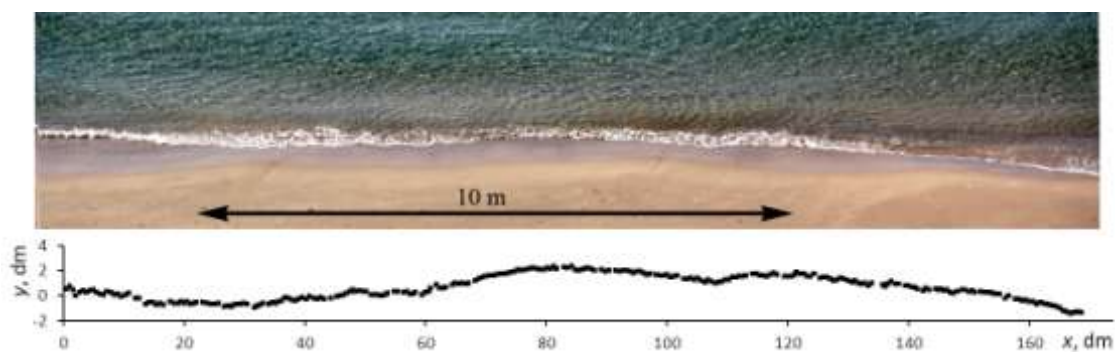

FIG. 2. Advancing contact line and its digitized presentation $y(x)$.

Characteristic lengths featuring the shape of the swash front were evaluated with the Fourier expansion of the swash front defined by the coordinates $y$ and $x$ as:

$$y(x) = \sum_m a_m e^{ik_n x}, \qquad k_m = 2\pi/\lambda_m, \qquad (2)$$

where $a_m$ are expansion coefficients corresponding to wavelengths $\lambda_m$. Figure 3 presents modules of the expansion coefficients $a_m$ depending on the wavelengths $\lambda_m$. According to Fig. 3, the dominant contribution to the roughness of swash front is given by the wavelengths in the range of 8 m.

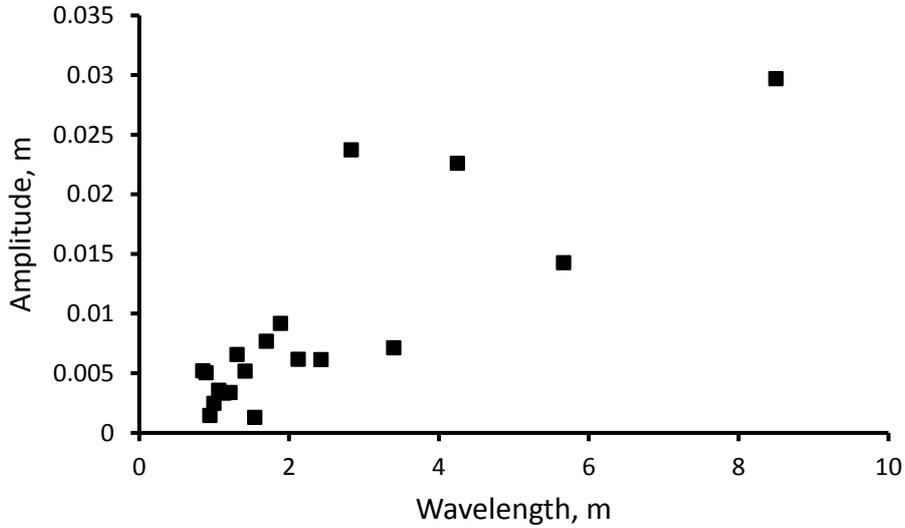

FIG. 3. Typical Fourier spectra of receding swash front line.

It is noteworthy that the dominant wavelengths established in our experiments coincide with "typical beach cusps" lengths [2, 29].

Roughness exponents of swash front lines were calculated according to the method described earlier [28, 15, 16]. A part of swash front with a total length $D$ =1700 cm was expressed by $n$ coordinates $y_i$ (width) and $x_i$ (length) differing by the value $\Delta x = 6.3$ cm. Segments with the length $L$ including $n(L)$ points were chosen, and for each segment its rms width $\sigma(L)$ (roughness) was calculated according to:

$$\sigma^2(L) = \frac{1}{L} \sum_{i=1}^{n(L)} [y_i - \bar{y}(L)]^2 \Delta x, \qquad (2)$$

$$\bar{y}(L) = \frac{1}{L} \sum_{i=1}^{n(L)} y_i \Delta x. \qquad (3)$$

After that, the average roughness $w(L)$ on ensemble of segments with the same $L$ centered in successive points of swash front $x_{0j}$ were calculated according to the equation

$$w(L) = \left( \frac{1}{D-L} \sum_j \sigma^2(L, x_{0j}) \Delta x_{0j} \right)^{0.5}, \quad (4)$$

$\Delta x_{0j}$ being the distance between the centers. Finally, the roughness of the swash edge line was characterized by its average rms width $w$ dependent on the length $L$. Dependences $w(L)$ presented in Fig. 4 are fine approximated with the scaling law $w(L) \propto L^\zeta$. Value of exponent $\zeta$ for receding swash front line was 0.64±0.02, when in the case of advancing swash the value 0.73±0.03 was calculated. The value of the scaling exponent $\zeta$ established for the receding phase of the swash are very close to the value of the exponent $\zeta=0.60\pm0.05$, established for the roughness of the triple line for water droplets deposited in rough surfaces [22]. The same exponent arises for a crack propagation front in a heterogeneous Plexiglas [27], and for the motion of a magnetic domain wall [12].

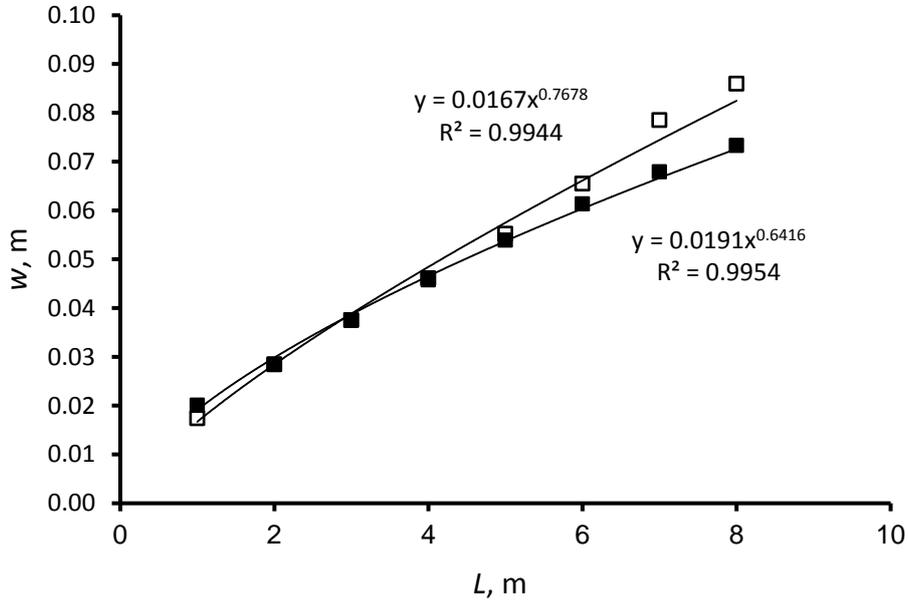

FIG. 4. Typical graphs of average roughness $w$ of the contact line versus length $L$ for advancing (open squares) and receding (solid squares) lines.

This coincidence deserves discussion. It may be suggested, that the swash front, the triple line and a crack propagation front may be described as randomly disturbed elastic strings [27, 30–31]. Rosso and Krauth demonstrated, that the anharmonic elastic potential of the randomly disturbed string, containing the second and fourth powers of the string displacement, leads to almost the same value of $\zeta$ equal to 0.63 [30]. Moreover, the same exponent appears in models based on directed percolation [32]. In these models, the mobile interface is stopped by a direct percolation cluster of pinning sites [32]. We conclude that the fine structure of the swash edge line is reasonably described by models based on the anharmonic elastic potential of the randomly disturbed strings. At the same time the scaling law appropriate for advancing swash is featured by the higher value of $\zeta = 0.73 \pm 0.03$. This discrepancy is plausible to relate to the dependence of advancing swash on the hydrodynamics of running waves.